\documentclass[aps,prl,twocolumn,superscriptaddress]{revtex4-2}
\newcommand{\fscale}{1.}

\newcommand{\ksig}{K^0_L\rightarrow \gamma X}
\newcommand{\kgg}{K^0_L\rightarrow \gamma \gamma}
\newcommand{\kpnn}{K^0_L\rightarrow\pi^0 \nu \bar{\nu}}
\newcommand{\kpp}{K^0_L\rightarrow\pi^0\pi^0}
\newcommand{\kppp}{K^0_L\rightarrow\pi^0\pi^0\pi^0}

\newcommand{\KLz}{K^0_L}

\usepackage{amsmath}
\usepackage{comment}
\usepackage{graphicx}
\usepackage{dcolumn}
\usepackage{bm}
\usepackage{color}

\usepackage{hyperref}
\usepackage{threeparttable}
\usepackage{lineno}
\usepackage{soul}

\begin{document}

\preprint{APS/123-QED}

\title{Search for Dark Particles in $\ksig$ at the KOTO Experiment}%


\newcommand{\AffChicago}{\affiliation{Enrico Fermi Institute, University of Chicago, Chicago, Illinois 60637, USA}}
\newcommand{\AffJeonbuk}{\affiliation{Division of Science Education, Jeonbuk National University, Jeonju 54896, Republic of Korea}}
\newcommand{\AffKorea}{\affiliation{Department of Physics, Korea University, Seoul 02841, Republic of Korea}}
\newcommand{\AffNTU}{\affiliation{Department of Physics, National Taiwan University, Taipei 10617, Taiwan, Republic of China}}
\newcommand{\AffKEK}{\affiliation{Institute of Particle and Nuclear Studies, High Energy Accelerator Research Organization (KEK), Tsukuba, Ibaraki 305-0801, Japan}}
\newcommand{\AffNDA}{\affiliation{Department of Applied Physics, National Defense Academy, Kanagawa 239-8686, Japan}}
\newcommand{\AffOsaka}{\affiliation{Department of Physics, Osaka University, Toyonaka, Osaka 560-0043, Japan}}
\newcommand{\AffYamagata}{\affiliation{Department of Physics, Yamagata University, Yamagata 990-8560, Japan}}
\newcommand{\AffSaga}{\affiliation{Department of Physics, Saga University, Saga 840-8502, Japan}}
\newcommand{\AffJPARC}{\affiliation{J-PARC Center, Tokai, Ibaraki 319-1195, Japan}}
\newcommand{\AffNKNU}{\affiliation{Department of Physics, National Kaohsiung Normal University, Kaohsiung 824, Taiwan}}
\newcommand{\AffNCUE}{\affiliation{Department of Physics, National Changhua University of Education, Changhua 50007, Taiwan}}

\newcommand{\MarkChiba}{\altaffiliation[Present address: ]{Department of Physics and The International Center for Hadron Astrophysics, Chiba University, Chiba 263-8522, Japan.}}

\newcommand{\MarkKEK}{\altaffiliation[Present address: ]{Institute of Particle and Nuclear Studies, High Energy Accelerator Research Organization (KEK), Tsukuba, Ibaraki 305-0801, Japan.}}

\newcommand{\MarkCERN}{\altaffiliation[Present address: ]{CERN, European Organization for Nuclear Research, CH-1211 Geneva, Switzerland.}}

\newcommand{\MarkDeceased}{\altaffiliation{Deceased.}}
\author{T.~Wu}\AffNTU
\author{Y.~C.~Tung}\AffNKNU
\author{Y.~B.~Hsiung}\AffNTU

\author{J.~K.~Ahn}\AffKorea
\author{M.~Gonzalez}\MarkCERN\AffOsaka
\author{E.~J.~Kim}\AffJeonbuk
\author{T.~K.~Komatsubara}\AffKEK\AffJPARC
\author{K.~Kotera}\AffOsaka
\author{S.~K.~Lee}\AffJeonbuk
\author{G.~Y.~Lim}\AffKEK\AffJPARC
\author{C.~Lin}\AffNCUE
\author{T.~Matsumura}\AffNDA
\author{H.~Nanjo}\AffOsaka
\author{Y.~Noichi}\AffOsaka
\author{T.~Nomura}\AffKEK\AffJPARC
\author{T.~Nunes}\AffOsaka
\author{K.~Ono}\AffOsaka
\author{J.~Redeker}\AffChicago
\author{N.~Shimizu}\MarkChiba\AffOsaka
\author{S.~Shinohara}\MarkKEK\AffOsaka
\author{K.~Shiomi}\AffKEK\AffJPARC
\author{R.~Shiraishi}\MarkKEK\AffOsaka
\author{Y.~Tajima}\AffYamagata
\author{Y.~W.~Wah}\AffChicago
\author{H.~Watanabe}\AffKEK\AffJPARC
\author{T.~Yamanaka}\AffOsaka
\author{H.~Y.~Yoshida}\AffYamagata
\collaboration{KOTO Collaboration} \noaffiliation


\begin{abstract}
We report a search for an invisible particle $X$ in the decay $\ksig$
($X \to \text{invisible}$), 
where $X$ can be interpreted as a massless or massive dark photon. 
No evidence for $X$ was found,
based on 13 candidate events consistent with a predicted background of 
$12.66 \pm 4.42_{\text{stat.}} \pm 2.13_{\text{syst.}}$ events.
Upper limits on the branching ratio of $\ksig$ were set for the $X$
mass range $0 \leq m_X \leq 425$ MeV/$c^2$.
For massless $X$, 
the upper limit was $3.4\times10^{-7}$ at the $90\%$ confidence level,
improving the previous indirect bound by over three orders of magnitude. 
This result yielded a new lower bound of 
$4.1\times10^6$~TeV 
on the probed effective mass scale governing 
flavor-changing dark-photon couplings,
representing approximately two orders of magnitude improvement on existing constraints.
For massive $X$, the upper limits in the searched mass region ranged 
from $\mathcal{O}(10^{-7})$ to $\mathcal{O}(10^{-3})$. 
\end{abstract}

\maketitle

\textit{Introduction~} The KOTO experiment at J-PARC~\cite{jparc} 
was designed for a dedicated search for
the rare decay $\kpnn$, a process that is challenging to identify due to the lack of kinematic constraints. 
To meet this challenge, 
together with the intense proton beam at J-PARC that 
provides a large number of kaons, 
KOTO employs a hermetic veto system with exceptional background-rejection capability.
These features have enabled KOTO to set the current world’s best upper limit on the branching ratio 
of $\kpnn$~\cite{kpnn}. 
The same experimental environment also provides a powerful opportunity to explore physics beyond the Standard Model (SM), including searches for hidden particles~\cite{kpnn}\cite{kxx}.

Here, we present a search for a hidden particle $X$ in the decay $K^0_L \rightarrow \gamma X$, 
where $X$ can be interpreted as a dark photon, 
the gauge boson of a new Abelian symmetry $U(1)_D$ invoked in extensions of the SM with dark sectors
~\cite{darkp}\cite{massless}. 
This decay provides one of the most sensitive probes
of dark photon couplings to SM particles 
at the fundamental quark level
via flavor-changing neutral current (FCNC) magnetic-dipole-type processes~\cite{emidio}\cite{jusak}.
We probed the mass range of $X$ from 0 to 425 MeV/$c^2$, under the assumption that $X$ decays invisibly. 
This region covers two theoretical scenarios: 
a massive dark photon ($A'$) 
arising from a spontaneously broken $U(1)_D$, or a massless dark photon ($\bar{\gamma}$) if the symmetry remains unbroken. 
In the massive case, 
$A'$ could couple to the SM particles 
through kinetic mixing, 
and the decay $K^0_L \to \gamma A'$ was discussed in~\cite{Smith}.
In the unbroken case, 
$\bar{\gamma}$ couples to SM particles 
only through higher-dimensional operators.
The leading contributions inducing the 
$s \to d \bar{\gamma}$ transition
arise from 
dimension-five FCNC magnetic-dipole-type operators, 
described by the effective Lagrangian
\begin{equation}
\mathcal{L}_{sd\bar{\gamma}} = \frac{1}{\Lambda_V} (\bar{s} \sigma_{\mu\nu} d) F^{\mu\nu} + \frac{1}{\Lambda_A} (\bar{s} \sigma_{\mu\nu} \gamma_5 d) F^{\mu\nu} + \text{H.c.}, 
\end{equation}
where $s$ and $d$ are the quark fields, 
$F^{\mu\nu}$ is the dark-photon field-strength tensor,
and $\Lambda_{V,A}$ are generally complex 
effective mass scales
encoding the heavy-messenger sector.
The branching ratio of $K^0_L \to \gamma \bar{\gamma}$ 
is directly controlled by the scales of $\Lambda_{V,A}$
and can be written numerically as
\begin{equation}
 \begin{gathered}
  \mathcal{B}( K^0_L \to \gamma \bar{\gamma}) \simeq \\
  5.74\times 10^{12} [ (\text{Re}~\Lambda_V^{-1})^2 
  + (\text{Im}~\Lambda_A^{-1})^2 ]~\text{GeV}^2~\text{\cite{jusak}}.
 \end{gathered}
 \label{eq:br_kgg}
\end{equation}
Using existing constraints on $\Lambda_{V,A}$, 
it has been estimated that the branching ratio of
$K^0_L \to \gamma \bar{\gamma}$ could be as large as
$1.2 \times 10^{-3}$~\cite{jusak}.
This article reports the first and high-sensitivity search for 
$K^0_L \to \gamma \bar{\gamma}$
that yields substantially stronger bounds
on the effective scales $\Lambda_{V,A}$.


\begin{figure*}[tb]
	\centering
	\includegraphics[width=0.9\linewidth]{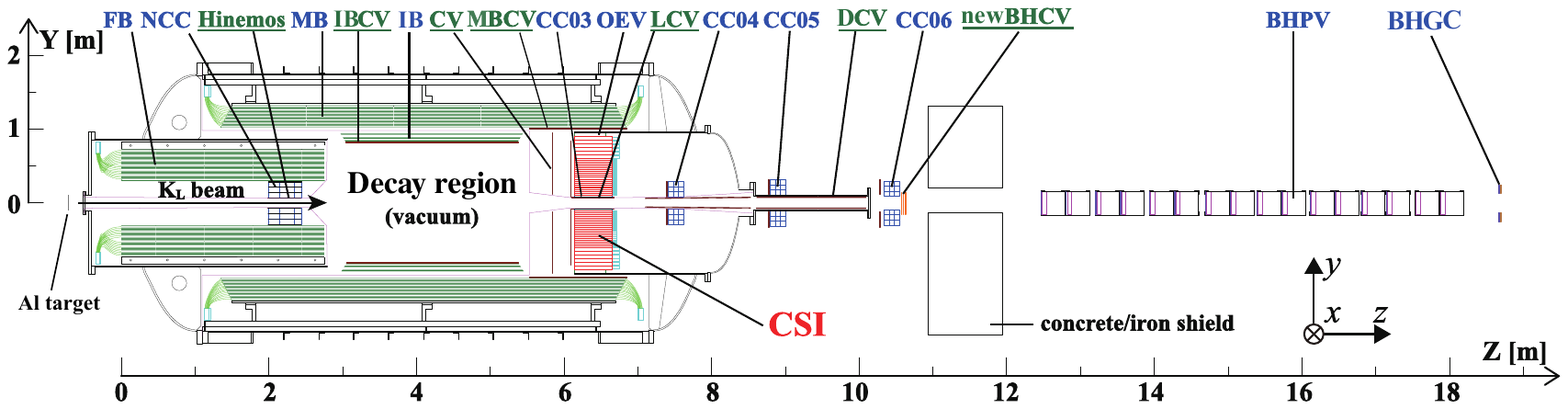}
	\caption{ 	
Cross-sectional view of the KOTO detector, with the beam entering from the left. The detector components with their names underlined represent charged-particle veto counters. The other components, except for CSI and Al target, 
serve as photon veto counters.
Al target is used only during dedicated neutron runs.
}
	\label{fig:KOTODetector}
\end{figure*}

\textit{KOTO apparatus~} Neutral kaons are produced by 30-GeV protons from the J-PARC Main Ring incident on a gold target~\cite{target} 
and are extracted at 16° to the proton beam using two collimators
with a sweeping magnet between them~\cite{beamline}. 
The resulting neutral beam
has a $K_L^0$ momentum distribution peaking at 1.4 GeV/$c$, a solid angle of 7.8 µsr, and a transverse size of $8\times8$ cm$^2$ at the exit of the downstream collimator.
The $K_L^0$s are transported to the KOTO detector located 21.5 m downstream of the target.
In addition, 
the neutral beam contains a substantial neutron flux, approximately 30 times larger than that of $K_L^0$ particles
for neutrons with momenta from 0.1~GeV/$c$ up to
several GeV/$c$.
The remaining beam components are dominated by photons.

\autoref{fig:KOTODetector} shows a cross-sectional
view of the KOTO detector, described in detail 
in~\cite{taku, csi, ib, cv}. 
At the downstream end of the decay volume, 
the CsI calorimeter (CSI)
measures the energies and positions of photons~\cite{csi}.
All other detector components serve as veto counters
to suppress backgrounds.
Some of these, 
hermetically surrounding the decay volume and CSI, 
include lead–scintillator barrel counters 
(FB, MB, and IB~\cite{ib}), 
collar counters (NCC and CC03–06), 
and charged-particle veto detectors 
(IBCV, CV~\cite{cv}, and LCV).
The downstream beam region also includes 
charged-particle veto counters (DCV and newBHCV) and 
photon veto counters (BHPV and BHGC).

\textit{Data collection~} For this analysis, data was collected during a dedicated 2-hour run in June 2020. 
The duration of this dedicated run was determined to balance the total run time against the statistical return from the background control samples.
With the J-PARC accelerator operating at 51~kW, a total of $5.03\times10^{16}$ protons on target (POT) were delivered, producing $1.29\times10^{10}$ $\KLz$ decays 
at the exit of the $K_L^0$ beam line.
Events for the $\ksig$ search were selected online using a two-level trigger system. 
The Level-1 (L1) trigger required the total energy in CSI to exceed 300 MeV with no coincident activity in the CV, IB, MB, NCC, or CC03–CC06 veto detectors. 
The Level-2 (L2) trigger further required exactly one 
photon in CSI. 
To study trigger effects and determine the $\KLz$ flux, 
an additional trigger with looser L1 veto conditions 
and no L2 selection was simultaneously employed
with a prescale factor of 13.

Separate neutron runs were conducted to collect beam neutron events for studying the neutron background in the $\ksig$ analysis.
During the neutron run, 
a 3-mm-thick aluminum (Al) target, shown in \autoref{fig:KOTODetector},
was inserted into the beam at 669~mm upstream of the detector entrance to scatter neutrons and enhance their interactions with CSI. 
Events were collected with the same L1 veto requirements as for $\ksig$, but with the L2 condition requiring two clusters in CSI.
These two-cluster events typically occur when a single neutron initially interacts in CSI, producing one cluster, and then scatters elsewhere within CSI, producing a second, spatially separated cluster.
The L2 two-cluster requirement was adopted 
because one-cluster neutron data was contaminated 
by $\kgg$ decays 
in which a photon escaped detection due to veto inefficiency.
Contamination from $\kgg$ events in the two-cluster sample was greatly suppressed by excluding events in which the center-of-energy of the two clusters was near the center of CSI and the opening angle between the position vectors of the two clusters on the CSI surface plane was close to $180^\circ$.


\textit{Event reconstruction and selection~} The particle $X$ and its decay products were assumed to be non-interacting with the KOTO detector; 
therefore, a $\KLz \to \gamma X$ event was characterized by a single photon in CSI with no coincident activity in any veto detector. 
The cluster produced by photon interactions in CSI was reconstructed by 
grouping neighboring crystals with energy deposits above 3~MeV within 
a 150-ns time window.
The resulting cluster provided measurements of the photon's energy ($E_\gamma$) and position $(x, y)$ on the upstream surface of CSI. 
To ensure full containment of the photon's energy,
the cluster position was required to lie within the fiducial region defined by $H_{XY} > 175$~mm and $\sqrt{x^2 + y^2} < 850$~mm, where $H_{XY}$ denoted the larger of $|x|$ or $|y|$. 
The lower bound on $H_{XY}$ also suppressed backgrounds from upstream kaon decays, in which a kaon decayed before entering the detector and one of its decay products 
hit the inner edge of CSI.
The photon energy was required to exceed 500~MeV.
The signal region was further 
defined by requiring
$900 < E_\gamma < 3000$~MeV and $325 < H_{XY} < 850$~mm.
The broader region, $800 < E_\gamma < 3000$~MeV and $300 < H_{XY} < 850$~mm, which marginally covered the signal region to avoid edge-tuning bias, 
was kept blinded until all event selection criteria were finalized.

To ensure that no particles were present other than a single photon in CSI, 
events were discarded if in-time energy deposits in the veto detectors exceeded the following thresholds: 1~MeV for FB, MB, IB, NCC, MBCV, IBCV, and OEV; 3~MeV for CC03–CC06; and 0.2~MeV for CV; Furthermore, 
BHPV was required to have less than three consecutive modules 
with in-time energy deposits.

\begin{figure}[h!]
  \centering
  \includegraphics[width=\fscale\linewidth]{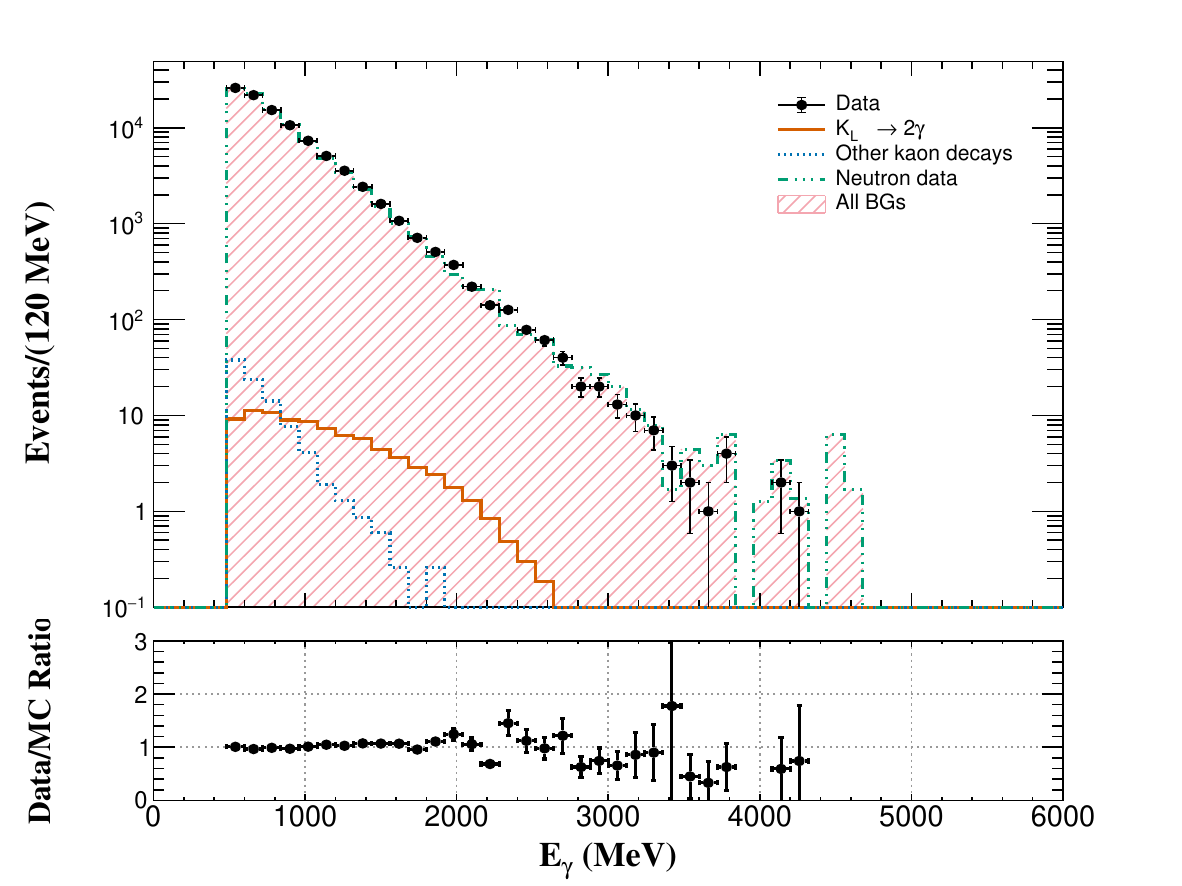}
  \includegraphics[width=\fscale\linewidth]{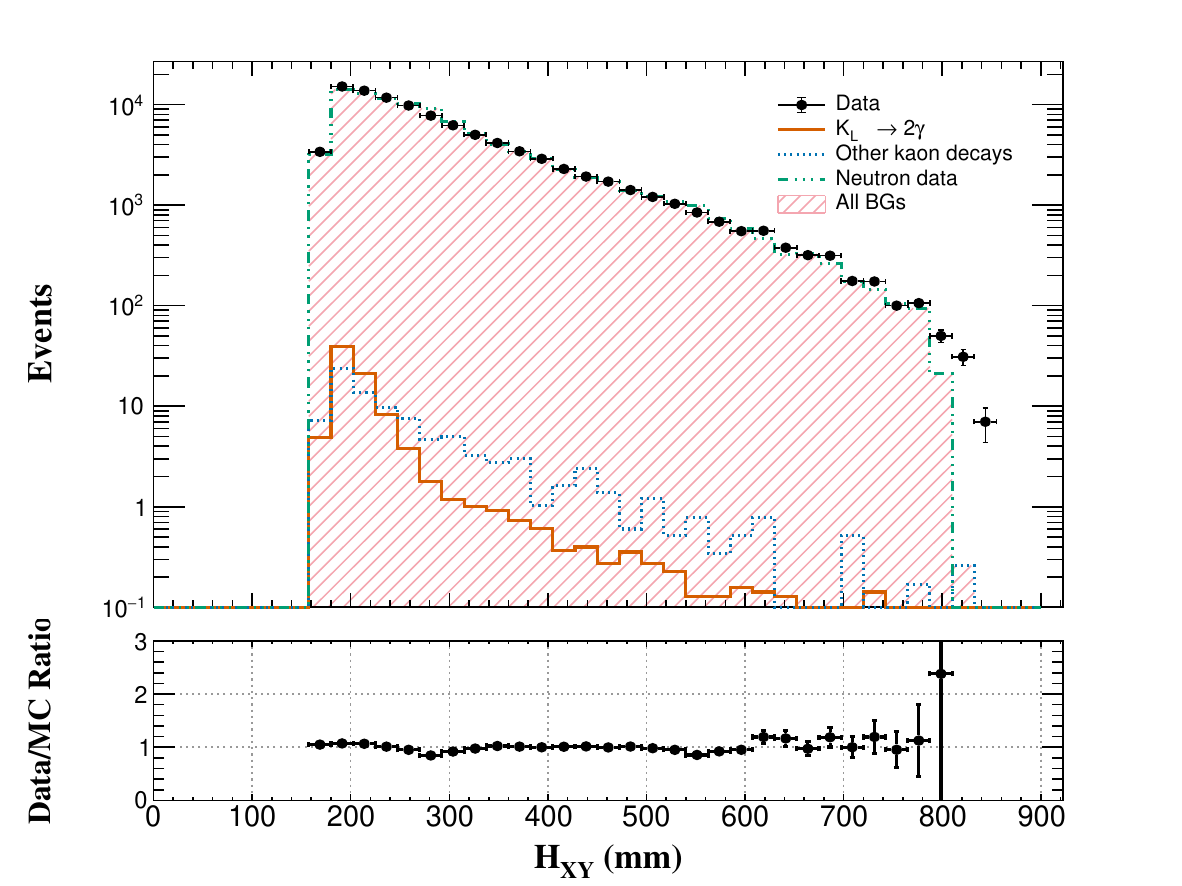}
  \caption{Distributions of $E_{\gamma}$ and $H_{XY}$ after requiring no in-time signal in veto counters. 
  Black dots represent the  data, while the hollow histograms show the contributions from each background source indicated in the figure. ``Other kaon decays'' include all $K_L^0$ backgrounds summarized in \autoref{tab:BGSummary}, excluding $\kgg$. 
The neutron data corresponds to samples collected in dedicated neutron runs.
The shaded histogram represents the sum of all backgrounds.
  }
  \label{fig:hxy}
\end{figure}

\textit{Background suppression~} Backgrounds for $\KLz \to \gamma X$ 
were classified into two main categories: 
neutron- and kaon-induced sources. 
The neutron-induced background was evaluated using data collected during dedicated neutron runs.
Contributions from various $K_L^0$ decay channels were estimated using {\scshape Geant4}-based Monte Carlo (MC) simulations~\cite{geant1}\cite{geant2}. 
The MC was further overlaid with accidental data collected simultaneously with physics data to reproduce actual beam activity and random electronic noise.
The agreement between the data and the modeled background spectra is shown in \autoref{fig:hxy}.

The dominant background arises from beam neutrons that acquire sufficient transverse momentum through 
elastic scattering at 
the surfaces of the beam-line collimators.
These neutrons can reach CSI 
through the central apertures or 
by traversing the inner edges of NCC 
without depositing significant energy (below 1 MeV) 
to trigger a veto, 
producing a single cluster 
that mimics the $\ksig$ signature.
This background was studied using two-cluster events collected during dedicated neutron runs, where
the higher-energy cluster, typically produced by the neutron’s initial interaction in CSI, 
was selected to model the neutron background in $\ksig$.
The selected neutron cluster was normalized to the $\ksig$ data using events 
identified as neutron-like by their cluster shapes (as described below)
in the region $800 < E_\gamma < 3000$~MeV and $175 < H_{XY} < 300$~mm. 
To account for differences in neutron cluster positions between two-cluster neutron data and one-cluster $\ksig$ data, the $H_{XY}$ distribution of the selected neutron cluster was weighted to match that of the $\ksig$ data after normalization,
without altering other measured properties of the neutron events.

The neutron background was suppressed using three techniques: Cluster Shape Discrimination (CSD), Pulse Shape Discrimination (PSD), and Shower Depth Measurement (SDM). 
CSD distinguished neutrons from photons based on differences in the cluster patterns produced by hadronic interactions of neutrons and electromagnetic interactions of photons. PSD exploited differences in the pulse shapes recorded in individual CSI channels to discriminate neutrons from photons. 
SDM measured the shower depth along the $z$-direction in CSI by evaluating the pulse timing difference
between the upstream MPPC and downstream PMT at the two ends of each CsI crystal.
Photon-induced electromagnetic showers typically developed near the upstream surface, whereas neutron-induced hadronic interactions 
could occur more uniformly throughout the crystal, 
as the interaction length of CsI 
is approximately $20$ times larger than
its radiation length.
Further details of CSD, PSD, and SDM are given 
in~\cite{csd}\cite{kotera}.
By combining these three techniques, the neutron background was suppressed 
by a factor of 560, resulting in 
$11.57 \pm 4.42_{\text{stat.}} \pm 2.13_{\text{syst.}}$ events in the signal region. 
The statistical uncertainty 
was dominated by the limited statistics of the neutron data sample, 
while the systematic uncertainty, estimated using neutron-like events for normalization, accounted for discrepancies in individual selection criteria (cuts) between $\ksig$ and neutron data.

The $\kgg$ decay was the most dominant background 
among all $K_L^0$ decays, 
as it shares a similar phase space with $\ksig$ and 
could mimic the signal if one of the final-state photons 
escaped detection. 
After applying all selection criteria, 
the number of $\kgg$ background events in the signal region was estimated to be 
$1.09 \pm 0.12_{\rm stat} \pm 0.05_{\rm syst}$.
The systematic uncertainty was evaluated using two-cluster 
$\kgg$ data and MC events, in which one of the two photons was randomly selected to compare their differences.
Backgrounds from other $K_L^0$ decay channels are summarized in \autoref{tab:BGSummary}.
Combining all background sources,
the total number of background events in the signal region was evaluated to be 
$12.66 \pm 4.42_{\text{stat.}} \pm 2.13_{\text{syst.}}$.

\begin{table}[!]
  \caption{Summary of background estimation.}
  \label{tab:BGSummary}
  \centering
  \begin{threeparttable}[h]
    \begin{tabular}{llc}
      \hline \hline
      Source & & Number of events\\
      \hline
      $K_L^0$ decay    & $\kgg$               & $1.09 \pm 0.12_{stat.}\pm 0.05_{syst.}$     \\
      & $\kppp$                  & $<0.18$ (90\% C.L.) \\
      & $\kpp$                  & $<0.05$ (90\% C.L.) \\
      & $K_L^0\to\pi^\pm e^\mp \nu_e$     & $<0.27$ (90\% C.L.) \\
      & $K_L^0\to\pi^\pm \mu^\mp \nu_\mu$ & $<0.25$ (90\% C.L.) \\
      & $K_L^0\to\pi^+ \pi^- \pi^0$       & $<0.17$ (90\% C.L.) \\
      Neutron                                                           &                                 & $11.57 \pm 4.42_{stat.}\pm 2.13_{syst.}$     \\
      \hline
      \multicolumn{2}{l}{Total (central value)} & $12.66 \pm 4.42_{stat.} \pm 2.13_{syst.}$                                      \\
      \hline \hline
    \end{tabular}
  \end{threeparttable}
\end{table}

\textit{Single event sensitivity and systematic uncertainties~} The Single Event Sensitivity (SES) was defined as the branching ratio corresponding to the observation of one signal event and was calculated as
\begin{align}\label{eq:SES-rephased}
  SES = \frac{A_\text{norm}}{A_\text{sig}} \times \frac{1}{N_\text{norm}} \times \mathcal{B}_{\text{norm}} \text{,}
\end{align}
where $A_\text{norm}$, $N_\text{norm}$, and $\mathcal{B}_{\text{norm}}$ 
denoted the acceptance, the number of observed events, and 
the nominal branching ratio of $\kppp$ from~\cite{pdg}, respectively.
From these quantities, 
the $K_L^0$ flux was obtained as
${N_\text{norm}}/({A_\text{norm}} \times {\mathcal{B}_\text{norm}}) = 
(1.29 \pm 0.02)\times 10^{10}$,
with $N_\text{norm}$ corrected for the online trigger prescale factor.
The signal acceptance $A_\text{sig}$,
which includes the $K_L^0$ decay probability within the KOTO detector,
was obtained from the $\ksig$ MC simulation and 
evaluated to be $(2.697 \pm 0.002)\times 10^{-3}$ for massless $X$,
yielding an SES of
$(2.87 \pm 0.05_{\rm stat} \pm 0.30_{\rm syst})\times 10^{-8}$.
For massive $X$, the sensitivity becomes worse with increasing $X$ mass due to lower $E_\gamma$, causing events to fall below the 900-MeV lower bound of the signal region.

\autoref{tab:summary-syst-ses} 
summarizes the systematic uncertainties in the SES calculation arising from
$\left(a\right)$ the veto cuts,
$\left(b\right)$ the kinematic cuts for $\kppp$,
$\left(c\right)$ the kinematic cuts for $\ksig$,
$\left(d\right)$ the neutron cuts for $\ksig$,
$\left(e\right)$ the $K_L^0$ momentum spectrum, 
$\left(f\right)$ the trigger effects, and
$\left(g\right)$ Branching ratio of $\kppp$.
Uncertainties $\left(a\right)$, $\left(b\right)$, and $\left(c\right)$ were derived from discrepancies in acceptance between data and MC for individual selection cuts, with percentage differences combined in quadrature. 
The dominant contribution came from source $\left(a\right)$ ($6.6\%$).
For $\kppp$, the uncertainty was obtained by direct comparison of data and MC, while for $\ksig$ it was evaluated from differences in the photon spectrum of 
$\kgg$ data and MC. 
The uncertainty $\left(d\right)$ accounted for the photon acceptance of the aforementioned three neutron-selection cuts and 
was estimated similarly from comparisons between photons in $\kgg$ data and MC.
The uncertainty $\left(e\right)$ was estimated from the difference between the $K_L^0$ momentum spectra in data and MC. The trigger-related uncertainty $\left(f\right)$ was evaluated using data samples collected with looser L1 cuts and without L2 selections. 
The uncertainty $\left(g\right)$ in the branching ratio of $\kppp$ was taken from~\cite{pdg}.

\begin{table}[h!]
  \caption{Summary of Systematic Uncertainties in SES.}
  \centering
  \begin{tabular}{lc}
    \hline \hline
    Sources                      & Uncertainty \\
    \hline
    $\left(a\right)$ Veto cuts                    & $ 6.6\% $   \\
    $\left(b\right)$ Kinematic cuts for $\kppp$ & $ 3.3\% $   \\
    $\left(c\right)$ Kinematic cuts for $\ksig$   & $ 1.4\% $   \\
    $\left(d\right)$ Neutron cuts for $\ksig$     & $ 4.8\% $   \\
    $\left(e\right)$ $\KLz$ momentum spectrum     & $ 0.9\% $   \\
    $\left(f\right)$ Trigger effect               & $ 1.6\% $   \\
    $\left(g\right)$ Branching ratio of $\kppp$ & $ 0.6\% $   \\
    \hline
    Total                        & $ 9.1\% $   \\
    \hline \hline
  \end{tabular}
  \label{tab:summary-syst-ses}
\end{table}

\begin{figure}[htb!]
  \centering
  \includegraphics[width=1\linewidth]{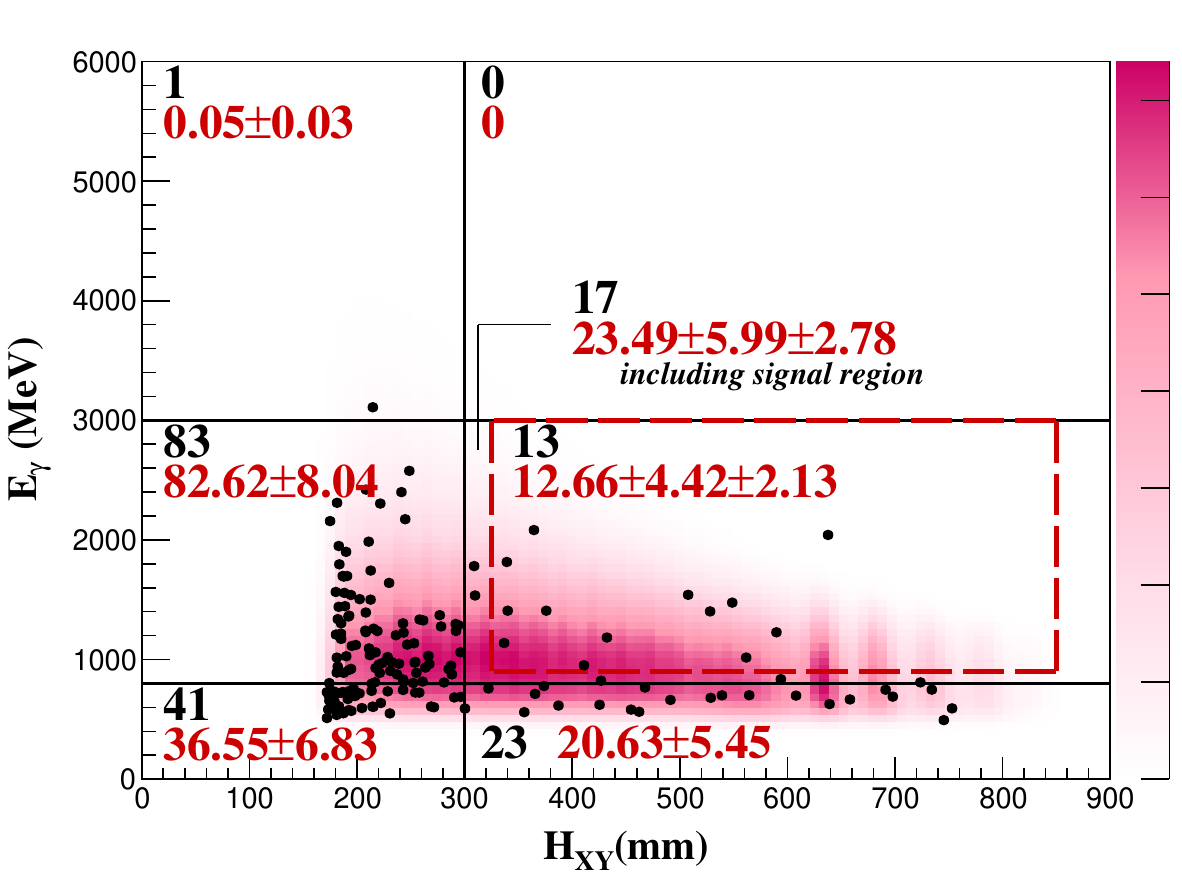}
  \caption{Distribution of $E_\gamma$ versus $H_{XY}$ after all selection criteria for $\ksig$. The red rectangle indicates the signal region. Black dots are the observed events, and the shaded contour shows the $\ksig$ MC distribution for massless $X$. 
Numbers in black (red) represent the observed (expected) event counts.}
  \label{fig:openbox}
\end{figure}

\begin{figure}[htb]
  \centering
  \includegraphics[width=1\linewidth]{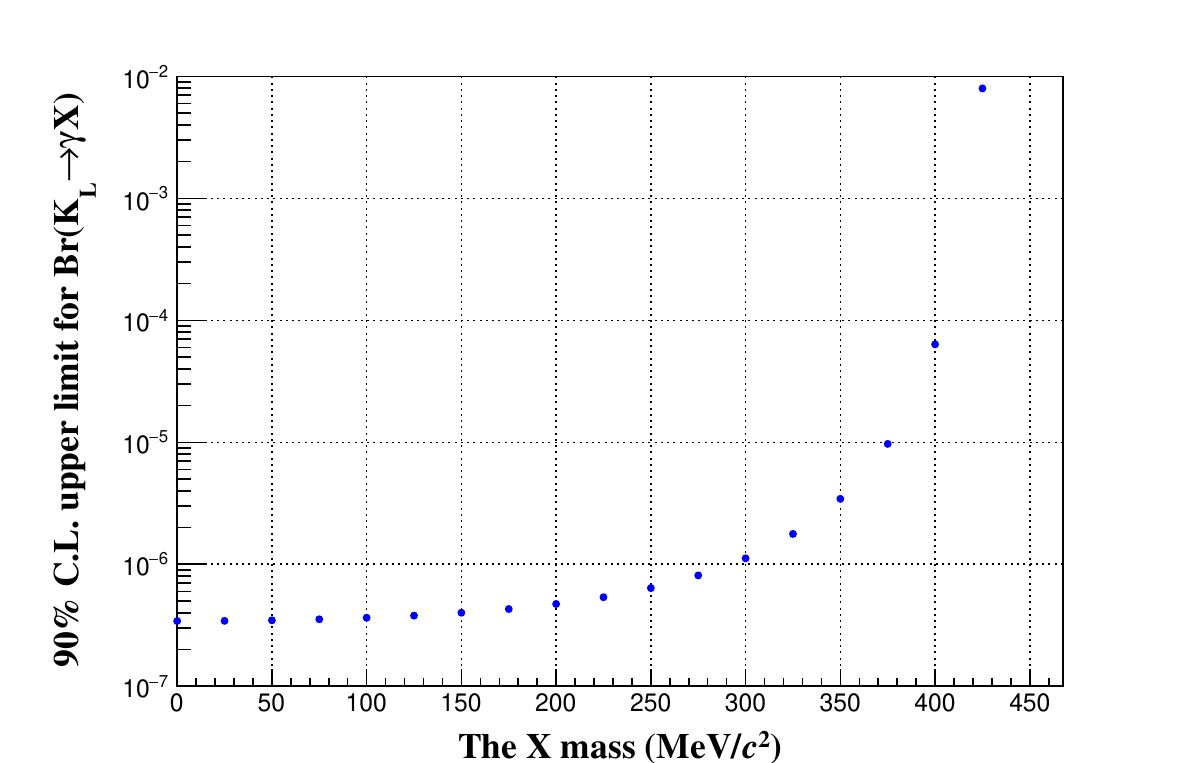}
  \caption{ Upper limit at the 90\% C.L. on the branching fraction of the 
  $\ksig$ decay as a function of the $X$ mass.}
  \label{fig:massive}
\end{figure}

\textit{Conclusions~} 
After finalizing the selection criteria, the signal region was unblinded, revealing 13 events in the signal region, as shown in \autoref{fig:openbox}. 
This observation was consistent with the expected background of 
$12.66 \pm 4.42_{\text{stat.}} \pm 2.13_{\text{syst.}}$, and no evidence for $\ksig$ was found. 
With the uncertainties in the background estimation incorporated,
an upper limit on the number of $\ksig$ candidates 
at the $90\%$ confidence level (C.L.) was
determined to be 11.9 events using the Feldman-Cousins method~\cite{Feldman1998}. 
This result was primarily limited by uncertainties in the background estimation; therefore, collecting a substantially larger physics dataset would not have meaningfully improved the final upper limit.
Upper limits on the branching ratio of $\ksig$ were derived for different masses of $X$ in the range 0--425 MeV/$c^2$. 
For massless $X$, 
the upper limit on the branching ratio was set to be 
$3.4 \times 10^{-7}$ at the $90\%$ C.L.,
while for massive $X$,
the $90\%$ C.L. upper limits ranged from 
$3.4\times10^{-7}$ at $m_X=25~\text{MeV}/c^2$ to $8.0\times10^{-3}$ 
at $m_X=425~\text{MeV}/c^2$, 
as shown in \autoref{fig:massive}.

Using Eq.~\eqref{eq:br_kgg}, the limit 
$\mathcal{B}(K_L^0\to\gamma\bar{\gamma}) < 3.4 \times 10^{-7}$ implies 
\begin{equation}
  (\text{Re}~\Lambda_V^{-1})^2
	+  
  (\text{Im}~\Lambda_A^{-1})^2	< 5.9\times 10^{-20}
  ~\text{GeV}^{-2}.	
\end{equation}
This bound applies specifically to the CP-allowed 
combination 
$(\mathrm{Re}\,\Lambda_V^{-1})^2 + (\mathrm{Im}\,\Lambda_A^{-1})^2$ 
that enters the $K_L^0 \to \gamma \bar{\gamma}$ decay rate. 
The orthogonal components, 
$\mathrm{Im}\,\Lambda_V^{-1}$ and 
$\mathrm{Re}\,\Lambda_A^{-1}$, 
do not contribute to this decay 
and are instead probed by complementary processes, 
such as BESIII constraints from
$c \to u \bar{\gamma}$ transitions~\cite{bes3}.
Assuming single-operator dominance, i.e., that either the vector or the axial operator contributes at a time, 
this corresponds to lower limits 
on the heavy-messenger scale of
$|\Lambda_V| \simeq |\Lambda_A| \gtrsim 4.1 \times 10^6~\text{TeV}$.
Compared with the previous indirect bound,
$\mathcal{B}(K_L^0\to\gamma\bar{\gamma}) 
\lesssim 1.2 \times 10^{-3}$,
the present result improves the upper limit by 
more than three orders of magnitude, 
corresponding to an increase of 
approximately two orders of magnitude 
in the probed heavy-messenger scale.
These are the most stringent constraints to date on 
flavor-changing dipole couplings between quarks and a dark photon, significantly restricting dark-sector scenarios 
with invisible final states.

We thank Dr. Jusak Tandean for his valuable discussions and theoretical insights on this study. We would like to express our gratitude to all members of the J-PARC Accelerator and Hadron Experimental Facility groups for their support. We also thank the KEK Computing Research Center for KEKCC, the National Institute of Informatics for SINET4, and the University of Chicago Computational Institute for the GPU farm. 
This material is based upon work supported by the Ministry of Education, Culture, Sports, Science, and Technology (MEXT) of Japan and the Japan Society for the Promotion of Science (JSPS) under KAKENHI Grant Numbers JP16H06343 and JP21H04995 and through the Japan-U.S. Cooperative Research Program in High Energy Physics; 
the U.S. Department of Energy, Office of Science, Office of High Energy Physics, under Awards No. DE-SC0009798; 
the National Science and Technology Council (NSTC) and Ministry of Education (MOE) in Taiwan, under Grant Numbers 
NSTC-108-2112-M-002-001, 
NSTC-109-2112-M-002-021, 
NSTC-109-2112-M-002-013,
NSTC-110-2112-M-002-020, MOE-109L892105, and MOE-110L890205 through National Taiwan University; 
the National Research Foundation of Korea under Grant Numbers 2020R1A3B2079993, RS-2022-NR070836, and RS-2025-00556834.

\nocite{*}

\bibliography{apssamp}

\end{document}